\documentstyle[preprint,aps,floats]{revtex}

\tightenlines

\begin{document}

\renewcommand{\thefootnote}{\fnsymbol{footnote}}

\begin{flushright}
BI-TP 97/53, hep-ph/9711487\\
to appear in 15 March 1998 issue of Phys. Rev. D (Rapid Comm.)
\end{flushright}

\vspace{0.5cm}

\centerline{{\large \bf
Improvement of the method of diagonal Pad\'e approximants}}
\centerline{{\large \bf
for perturbative series in gauge theories}}

\vspace{1.cm}

\centerline{ G.~Cveti\v c\footnote[1]{e-mail:
cvetic@physik.uni-bielefeld.de; 
or: cvetic@doom.physik.uni-dortmund.de}}

\centerline{{\it  
Department of Physics, Universit\"at Bielefeld,
33501 Bielefeld, Germany}}

\renewcommand{\thefootnote}{\arabic{footnote}}

\begin{abstract}

Recently it has been pointed out that diagonal
Pad\'e approximants to truncated perturbative series in
gauge theories have the remarkable property of being
independent of the choice of the renormalization scale
as long as the gauge coupling parameter ${\alpha}(p^2)$
is taken to evolve according to the one-loop renormalization
group equation -- i.e., in the large-${\beta}_0$
approximation. In this letter we propose and describe
an improvement to the method of diagonal Pad\'e approximants.
The improved method results in approximants which are independent 
of the chosen renormalization scale when ${\alpha}(p^2)$
evolves at any chosen (in principle arbitrary) loop-level.\\
PACS number(s): 11.10.Hi, 11.80.Fv, 12.38.Bx, 12.38.Cy

\end{abstract}

\setcounter{equation}{0}
\newpage

Pad\'e approximants (PA's), either diagonal or nondiagonal,
when applied to any truncated perturbative series (TPS), 
possess by construction the same formal accuracy\footnote{
We refer to Ref.~\cite{BakerMorris} (Part I) for
the basic theory of PA's.}
as the original TPS. This means that, when expanding the
applied PA in a power series of the expansion parameter of the TPS,
we reproduce all the coefficients of the TPS. PA's,
being rational fractions, additionally
act as a kind of analytical continuation of the
TPS and thus often represent a substantial improvement
of the results deduced directly from the TPS. 
This had been the main motivation for applying PA's to TPS's
in gauge theories such as QCD \cite{SEK}.\footnote{
For a review of methods of dealing with
power expansions in quantum field theory, see Ref.~\cite{Fischer}.}

Recently it has been noted \cite{Gardi} that the 
{\em diagonal} Pad\'e approximants (dPA's),
when applied to TPS's in gauge theories, have
in addition the remarkable property of being independent
of the choice of the renormalization scale (RScl) if
the gauge coupling parameter ${\alpha}(p^2)$ is taken
to evolve according to its one-loop renormalization group equation:
${\alpha}(p^2)\!=\!{\alpha}(q^2)/[1\!+\!{\beta}_0 \ln(p^2/q^2)
{\alpha}(q^2)]$. This is the direct consequence
of the mathematical property of dPA's that they are invariant
under the homographic transformations of the
argument (Ref.~\cite{BakerMorris}, Part I):
$z\!\mapsto\!az/(1+bz)$. Since a full observable (formally
an infinite perturbation series) must be RScl--independent,
the mentioned property of dPA's suggests that the dPA for
an available TPS of an observable in a gauge theory 
(approximately) sums
up a substantial set of diagrams and thus represents
a very reasonable resummation method there. The authors
of \cite{Brodskyetal} further investigated this dPA method
and showed that the resummed diagrams represent systematic
approximations to the Neubert's \cite{Neubertetal} concept of the
distribution of momentum flow through a bubble-dressed gluon
propagator. The authors of \cite{Brodskyetal} pointed out the
need for further improvements of the method, in particular
to obtain RScl-independence beyond the 
large-${\beta}_0$ (one-loop flow) limit.

We present here an algorithm which improves the dPA method
in this sense -- by constructing approximants
which are RScl-independent at {\em any} chosen loop-level
of evolution of ${\alpha}(p^2)$ and which, 
when expanded in power series of ${\alpha}$,
give the same formal accuracy as the TPS's to which
they are applied.

A generic observable $S$ in a gauge theory (e.g., QED or QCD)
can in general be redefined so that it has the following
form as a formal perturbation series:
\begin{equation}
S \equiv a(q^2) f(q^2) = a(q^2) \left[
1 + r_1(q^2) a(q^2) + r_2(q^2) a^2(q^2) + \cdots 
+ r_n(q^2) a^n(q^2) + \cdots \right] \ .
\label{S}
\end{equation}
Here, $a(q^2)\!\equiv\!{\alpha}(q^2)/\pi$ and $q^2$ is a
chosen renormalization scale (RScl). The full series (\ref{S}) is
of course independent of $q^2$. In practice we have only a limited
number of coefficients $r_j(q^2)$ available ($i\!=\!1,\!\ldots,\!n$),
i.e., we know only a truncated perturbation series (TPS)
\begin{equation}
S_n(q^2) \equiv a(q^2) f^{(n)}(q^2) = a(q^2) \left[
1 + r_1(q^2) a(q^2) + r_2(q^2) a^2(q^2) + \cdots 
+ r_n(q^2) a^n(q^2) \right] \ .
\label{Sn}
\end{equation}
This TPS explicitly depends on the RScl $q^2$ -- changing $q^2$
changes the value of $S_n$ in general by a term $\sim\!a^{n+2}$.
The RScl-independence of (\ref{S}) and the RScl-dependence of
(\ref{Sn}) suggest that an approximant that has the same
formal accuracy and is RScl-invariant is a good candidate
to be closer than (\ref{Sn}) to the full answer (\ref{S}).

We will now construct such approximants.
The gauge coupling parameter 
$a(p^2)\!\equiv\!{\alpha}(p^2)/\pi$ evolves according to
the perturbative renormalization group equation (RGE)
\begin{equation}
\frac{d a(p^2)}{d \ln p^2} = 
- \sum_{j=0}^{\infty} {\beta}_j a^{j+2}(p^2) \ ,
\label{alphaRGE}
\end{equation}
where ${\beta}_j$ are constants if particle threshold effects
are ignored. They are positive in QCD and negative
in QED. Only a limited number of these perturbative
coefficients ${\beta}_j$ 
(${\beta}_0,\!\ldots,\!{\beta}_3$) are known in QCD
(cf.~\cite{QCDbeta}, in ${\overline {\mbox{MS}}}$ scheme) and in 
QED (cf.~\cite{QEDbeta}, in ${\overline {\mbox{MS}}}$, MOM and
in on-shell schemes). Hence, in practice the RGE (\ref{alphaRGE})
will always be truncated at some level.
We now introduce the ratio of gauge coupling parameters at
two different renormalization scales $p^2$ and $q^2$
\begin{equation}
k( a_q,u ) \equiv \frac{ a(p^2) }{ a(q^2) }
\qquad \mbox{where: } \quad a_q=a(q^2) \ , \ u=\ln(p^2/q^2) . \ 
\label{kdef}
\end{equation}
Formally expanding this function in powers of 
$u\!\equiv\!\ln(p^2/q^2)$ results in the following series:
\begin{equation}
k( a_q,u ) = 1 + \sum_{j=1}^{\infty} u^j k_j(a_q) \ ,
\qquad \mbox{where: } \ k_j(a_q) = \frac{1}{j!} 
\frac{\partial^j}{\partial u^j} k(a_q, u) {\big |}_{u=0} \ .
\label{kTaylor}
\end{equation}
We note that $k_j(a_q)\!\sim\!a^j(q^2)$ since RGE (\ref{alphaRGE})
gives the connection
\begin{equation}
k_j(a_q) = (-1)^j \beta_0^j a^j(q^2) + 
{\cal O}\left( a^{j+1}(q^2) \right) \ , \quad
k_0(a_q) = 1 \ ,
\label{kjsapprox}
\end{equation}
and the terms of higher orders can also be explicitly
obtained from RGE (\ref{alphaRGE}). 
At this point we rearrange the formal power series (\ref{S})
for $S/a(q^2)$ into a related series in $k_j(a(q^2))$
\begin{equation}
S \equiv a(q^2) f(q^2) = 
a(q^2) \left[ 1 + \sum_{j=1}^{\infty} f_j(q^2) 
k_j \left( a(q^2) \right) \right] \ .
\label{Sinkj}
\end{equation} 
We note that the coefficient $f_j(q^2)$ depends solely
on the first $j$ coefficients $r_1(q^2),\!\ldots,\!r_j(q^2)$
of the original series (\ref{S}), as implied by relations
(\ref{kjsapprox}). In addition, $f_j(q^2)$ depends on
the first $j$ coefficients 
${\beta}_0,\!\ldots,\!{\beta}_{j-1}$ of RGE (\ref{alphaRGE}). 
We then define the corresponding formal series ${\cal F}(q^2)$,
which is in powers of $(- {\beta}_0 a(q^2))$
\begin{equation}
a(q^2) {\cal F}(q^2) \equiv a(q^2) \left[
1 + \sum_{j=1}^{\infty} f_j(q^2) 
(- 1)^j {\beta}_0^j a^j(q^2) \right] \ .
\label{F1}
\end{equation} 
We construct for $a(q^2) {\cal F}(q^2)$ the
diagonal Pad\'e approximants (dPA's) with argument $a(q^2)$
\begin{eqnarray}
a(q^2) [ M-1 / M ]_{ {\cal F} }(q^2) &=&
a(q^2) \left[ 1 + \sum_{m=1}^{M-1} 
{\tilde a}_m(q^2) a^m(q^2) \right]
\left[ 1 + \sum_{n=1}^{M} {\tilde b}_n(q^2) a^n(q^2) \right]^{-1} \ ,
\label{PAF11}
\\
a(q^2) {\cal F}(q^2) &=&
a(q^2) [ M-1 / M ]_{ {\cal F} }(q^2) + 
{\cal O} \left( a^{2M+1}(q^2) \right) \ .
\label{PAF12}
\end{eqnarray}
The above dPA depends only on the first $(2M\!-\!1)$ 
coefficients $f_j(q^2)$ 
($j\!=\!1,\!\ldots,\!2M\!-\!1$), due to the standard
requirement (\ref{PAF12}).
Since, as mentioned earlier, the coefficient $f_j(q^2)$ 
is a unique function of only $r_1(q^2),\!\ldots,\!r_j(q^2)$, 
the above dPA depends only on the first $2M\!-\!1$ coefficients
$r_1(q^2),\!\dots,\!r_{2M-1}(q^2)$ of the original
series (\ref{S}), i.e., it is uniquely determined once
we have the TPS (\ref{Sn}) with $n\!=\!2M\!-\!1$ available.
Unless we have an exceptional situation when the denominator
in the dPA (\ref{PAF11}) has multiple zeros as polynomial
of $a(q^2)$, we can uniquely decompose this dPA into a sum of
simple fractions
\begin{equation}
a(q^2) [ M-1 / M ]_{ {\cal F} }(q^2) =
a(q^2) \sum_{i=1}^M \frac{ {\tilde \alpha}_i }
{ \left[ 1 + {\tilde u}_i(q^2) {\beta}_0 a(q^2) \right] } 
= \sum_{i=1}^M {\tilde \alpha}_i  \frac{ a(q^2) }
{ \left[ 1 + {\tilde u}_i(q^2) {\beta}_0 a(q^2) \right] } \ .
\label{PAF1decomp}
\end{equation}
Here, $[-{\tilde u}_i(q^2) {\beta}_0]^{-1}$ are the $M$ zeros of the
denominator of the dPA (\ref{PAF11}) which is regarded as
a polynomial of $a(q^2)$. The above expression (\ref{PAF1decomp}) 
is a weighed sum of the one-loop-evolved 
gauge coupling parameters\footnote{
Evolved from the RScl $q^2$ to $p_i^2$,
by the one-loop (``large-${\beta}_0$'')
version of RGE (\ref{alphaRGE}).} 
$a(p_i^2)$, with generally complex scales $p_i^2$ determined by
the relation ${\tilde u}_i(q^2)\!=\!\ln (p_i^2/q^2)$,
and with weights ${\tilde \alpha}_i$.
The approximant that we are looking for is then obtained 
by replacing in the above weighed sum the one-loop-evolved
gauge coupling parameters with those which evolve according
to the full RGE (\ref{alphaRGE})
\begin{equation}
a(q^2) G_{ f }^{ [M-1/M] }(q^2) \equiv 
a(q^2) \sum_{i=1}^{M} {\tilde \alpha}_i  
k \left( a(q^2),{\tilde u}_i \right) \ .
\label{dBGAdef}
\end{equation}
The functions $k( a(q^2), {\tilde u}_i)$ appearing here 
are defined via (\ref{kdef}) as ratios of 
gauge coupling parameters $a$ at the RScl $q^2$ and the
new scales $p_i^2\!=\!q^2 \exp [ {\tilde u}_i(q^2) ]$
\begin{equation}
k \left( a(q^2), {\tilde u}_i \right) = a(p_i^2)/a(q^2)
\qquad \mbox{where: } \ \ln ( p_i^2/q^2 ) =
{\tilde u}_i(q^2) \ .
\label{fullas}
\end{equation}
We stress that $a(p_i^2)\!\equiv\!{\alpha}(p_i^2)/{\pi}$ 
($i\!=\!1,\!\ldots,\!M$) are the gauge coupling parameters
evolved from the RScl $q^2$ to $p_i^2$ by the RGE (\ref{alphaRGE})
whose loop-level precision (i.e., the number
of included coefficients ${\beta}_j$) can be chosen 
as high as possible\footnote{In QED and QCD, this would mean
inclusion of ${\beta}_0,\!\ldots,\!{\beta}_3$ since these
coefficients are now available in certain schemes -- 
see earlier discussion.},
{\em independently\/} of the number $n\!=\!2M\!-\!1$ of the
coefficients of the available TPS $S_n(q^2)$ of Eq.~(\ref{Sn}).
By (\ref{fullas}), the obtained approximants (\ref{dBGAdef})
can be written in a somewhat more transparent form
\begin{equation}
a(q^2) G_{ f }^{ [M-1/M] }(q^2) \equiv 
\sum_{i=1}^M {\tilde \alpha}_i a(p_i^2) \qquad
\mbox{where: } \quad 
p_i^2 = q^2 \exp \left[ {\tilde u}_i (q^2) \right] \ .
\label{dBGAres}
\end{equation}
Function $k(a,u)$ in (\ref{dBGAdef}), 
which depends on two (in general complex) arguments, 
can be called the kernel of the above approximant. We will call
the above approximant the modified diagonal Baker-Gammel approximant
(modified dBGA) with kernel $k$, since there exists
a certain (but limited) similarity with the diagonal 
Baker-Gammel approximants as defined in 
Ref.~\cite{BakerMorris} (Part II, Sec.~1.2).
We emphasize again that this modified dBGA of order $2M\!-\!1$,
for observable $S$ of Eq.~(\ref{S}),
is uniquely determined by the first $2M\!-\!1$ coefficients
$r_1(q^2),\!\ldots,\!r_{2M-1}(q^2)$ of the perturbative
series (\ref{S}), i.e., by the TPS $S_{2M-1}$ of (\ref{Sn})
($n\!=\!2M\!-\!1$), and by the independently
chosen loop-level for the
evolution of $a(p^2)$ according to RGE (\ref{alphaRGE}). 

Having the value $a(q^2)$,
we should stress that exact values $a(p_i^2)$
can be obtained only if we evolve the RGE 
(\ref{alphaRGE}) {\em numerically\/}
from $u\!=\!\ln (q^2/q^2)\!=\!0$ to
$u\!=\!\ln(p_i^2/q^2)\!=\!{\tilde u}_i(q^2)$.
This numerical integration should be performed
with additional care when ${\tilde u}_i(q^2)$'s are
complex.
Approximate values for $a(p_i^2)$ can be obtained by
perturbatively expanding the solution $a(p_i^2)/a(q^2)$ in
powers of $a(q^2)$, but such expansion would be reasonable
only if the chosen RScl $q^2$ is not far away from the
scales $p_i^2$. In any case, the resulting approximant
(\ref{dBGAres}) does not represent an analytical
formula once we go beyond the large-${\beta}_0$
approximation.

It can be shown that this modified dBGA (\ref{dBGAres}) of order
$2M\!-\!1$, for observable $S$ of Eq.~(\ref{S}), fulfils the
two requirements that we wanted to achieve:
\begin{enumerate}
\item
It has the same formal accuracy as the TPS $S_{2M-1}(q^2)$ of
(\ref{Sn}):
\begin{equation}
S = a(q^2) G_{ f }^{ [M-1/M] }(q^2) + 
{\cal O} \left( a^{2M+1}(q^2) \right) \ .
\label{Theor1}
\end{equation} 
\item
It is fully invariant under the change of the renormalization
scale $q^2$. In fact, the weights ${\tilde \alpha}_i$ and the
scales $p_i^2\!=\!q^2 \exp[ {\tilde u}_i(q^2) ]$
are separately independent of the chosen renormalization scale $q^2$.
\end{enumerate}
The formal proofs of these two statements will be given in a
longer paper \cite{Cvetic}. Furthermore, also the discussion
of similarities and differences between the presented modified dBGA's
(\ref{dBGAres}) and the usual dBGA's of Ref.~\cite{BakerMorris}
will be given in that longer paper.

Within the presented algorithm,
the case of one-loop evolution of ${\alpha}(p^2)$
(the large-${\beta}_0$ approximation)
means: $k(a,u)\!=\!1/(1\!+\!{\beta}_0 u \ a)$, and 
$k_j(a)\!=\!{\beta}_0^j (-a)^j$, when using notation
of Eqs.~(\ref{kdef})--(\ref{kjsapprox}). 
Therefore, in the one-loop
case, expansion (\ref{Sinkj}) for $f(q^2)\!\equiv\!S/a(q^2)$ 
and (\ref{F1}) for ${\cal F}(q^2)$ are identical. The modified
dBGA (\ref{dBGAdef}) is in this case reduced to the usual dPA
(\ref{PAF1decomp}).

One may worry what happens when the parameters
${\tilde \alpha}_i$ and ${\tilde u}_i(q^2)$ in the
modified dBGA (\ref{dBGAdef})-(\ref{dBGAres}) 
are not simultaneously real.
In that case, the modified dBGA could be complex.
Since relation (\ref{Theor1}) and the RScl-invariance
are valid for the entire modified dBGA's, they are
valid separately for their real
and imaginary parts. The observable $S$
is real, so we then just take the real part of expression
(\ref{dBGAres}). Since $a(q^2)$ and $S$ are real,
relation (\ref{Theor1}) implies that the imaginary part
of the modified dBGA $a(q^2) G_{ f }^{ [M-1/M] }(q^2)$
must be $\sim\!a^{2M+1}(q^2)$ or even less.

It may be helpful not to remain at this rather abstract level, 
but to write more explicit formulas for dBGA's 
(\ref{dBGAdef})-(\ref{dBGAres}) in the practically interesting
cases of $M\!=\!1$ and $M\!=\!2$.

In the case $M\!=\!1$ ($n\!\equiv\!2M\!-\!1\!=\!1$)
the method gives the same result 
as the effective charge (ECH) method \cite{Grunberg},
and the Brodsky-Lepage-Mackenzie (BLM) method \cite{BLM}
in the large-${\beta}_0$ approximation
\begin{eqnarray}
S_1(q^2) \equiv a(q^2) f^{(1)}(q^2) &=& 
a(q^2) \left[ 1+r_1(q^2) a(q^2) \right]
\qquad  \Rightarrow  
\nonumber\\
a(q^2) G_{ f }^{[0/1]}(q^2) &=& 
a(Q^2)  \quad \mbox{where: } \ \
Q^2 = q^2 \exp \left[ -r_1(q^2)/{\beta}_0 \right] \ .
\label{BLMrel}
\end{eqnarray}
Here, $a(p^2)\!\equiv\!{\alpha}(p^2)/\pi$ evolves according to
RGE (\ref{alphaRGE}) where the chosen loop-level is arbitrary. 
It is straightforward
to check directly that $a(Q^2)$ is RScl-invariant and that
(\ref{Theor1}) is satisfied.

In the case $M\!=\!2$ ($n\!\equiv\!2M\!-\!1\!=\!3$),
parameters of the dBGA (\ref{dBGAdef})-(\ref{dBGAres})
can also be obtained in
a straightforward, although algebraically more involved, manner.
RGE (\ref{alphaRGE}) implies
\begin{eqnarray}
k_1(a) & = & - {\beta}_0 a - {\beta}_1 a^2 - {\beta}_2 a^3 
- \ldots \ ,
\label{k1}
\\
k_2(a) &=& + {\beta}_0^2 a^2 + (5/2) {\beta}_0 {\beta}_1 a^3 +
\ldots \ , \qquad k_3(a) = - {\beta}_0^3 a^3 - \ldots \ .
\label{k2k3}
\end{eqnarray}
Here we use the short-hand notation 
$a\!\equiv\!a(q^2)\!\equiv\!{\alpha}(q^2)/\pi$.
Inverting relations (\ref{k1})--(\ref{k2k3}) yields expressions
for $a,\!a^2$ and $a^3$ in terms of $k_1(a),\!k_2(a)$
and $k_3(a)$. We insert then these expressions into the truncated series
$S_3(q^2)$ of Eq.~(\ref{Sn}) (if it is available) and thus 
obtain the first three coefficients of the rearranged truncated series 
for $S_3(q^2)$ of the form (\ref{Sinkj})
\begin{eqnarray}
f_1(q^2) &=& - \frac{r_1(q^2) }{ {\beta}_0 } \ ,
\qquad
f_2(q^2) = - \frac{ {\beta}_1 }{ {\beta}_0^3 } r_1(q^2)
+ \frac{1}{ {\beta}_0^2 } r_2(q^2) \ ,
\label{f11f12}
\\
f_3(q^2) &=& \left( - \frac{ 5 {\beta}_1^2 }{ 2 {\beta}_0^5 }
+ \frac{ {\beta}_2 }{ {\beta}_0^4 } \right)  r_1(q^2)
+ \frac{ 5 {\beta}_1 }{ 2 {\beta}_0^4 } r_2(q^2)
- \frac{1}{ {\beta}_0^3 } r_3(q^2) \ .
\label{f13}
\end{eqnarray}
With these coefficients, we form the truncated series
for $a(q^2) {\cal F}(q^2)$ (\ref{F1}), and the dPA
$a(q^2) [1/2]_{ {\cal F} }(q^2)$ (\ref{PAF12}) in the form
(\ref{PAF1decomp}). We then obtain expressions for
parameters ${\tilde u}_i(q^2)$ and ${\tilde \alpha}_i$ 
($i\!=\!1,\!M$) for the case of $M\!=\!2$
\begin{eqnarray}
{\tilde u}_{2,1} &=& 
\frac{\left[ (f_3-f_1 f_2) \pm \sqrt{ {\rm det} } \right]}{
\left[ 2 (f_2 - f_1^2) \right]} \ , \quad
{\tilde \alpha}_1 = \frac{( {\tilde u}_2 - f_1 )}{
( {\tilde u}_2 - {\tilde u}_1 )} \ , \quad 
{\tilde \alpha}_2 = 1 - {\tilde \alpha}_1 \ .
\label{tildeu21}
\\
\mbox{where: } \
{\rm det} &=& \left[ f_3 + f_1 (2 f_1^2 - 3 f_2) \right]^2 +
4 (f_2 - f_1^2)^3 \ .
\label{det}
\end{eqnarray}
The plus sign in (\ref{tildeu21}) corresponds to
${\tilde u}_2(q^2)$. For simplicity, we omitted
notation of the RScl-dependence in the coefficients 
$f_i(q^2)$ and ${\tilde u}_i(q^2)$. Of course,
expressions (\ref{f11f12})-(\ref{f13}) should be
inserted into (\ref{tildeu21})-(\ref{det}) in order
to obtain these parameters explicitly in terms of the
original coefficients $r_i(q^2)$ ($i\!=1,\!2,\!3$). 
When we insert these obtained parameters
into the dBGA expression (\ref{dBGAres}) ($M\!=\!2$), we get the
RScl-invariant approximation to $S_3(q^2)$, with
the RScl-invariant parameters ${\tilde \alpha}_i$ and
$p_i^2\!=\!q^2 \exp [ {\tilde u}_i(q^2) ]$ ($i\!=\!1,\!2$) 
explicitly dependent on the
original coefficients $r_i(q^2)$ ($i\!=1,\!2,\!3$) and
on the RGE coefficients ${\beta}_j$ ($j\!=\!0,\!1,\!2$)
of Eq.~(\ref{alphaRGE}). Although the parameters
obtained above for the case of $M\!=\!2$
contain dependence on the first three 
RGE beta-coefficients (${\beta}_0,\!{\beta}_1,\!{\beta}_2$),
we should emphasize that the evolution of the gauge
coupling parameters $a(p_i^2)$ appearing\footnote{
RGE evolution from RScl $q^2$ to the (possibly complex) 
scale $p_i^2$ is meant here.} 
in the dBGA (\ref{dBGAres}) can be
governed by the RGE (\ref{alphaRGE}) with a higher
chosen loop-level accuracy, e.g., by inclusion of
${\beta}_3$ there. On the other hand, {\em at least\/}
the first three coefficients 
(${\beta}_0,\!{\beta}_1,\!{\beta}_2$) should be
taken into account in the RGE evolution of $a(p^2)$
from the RScl $q^2$ to $p_i^2$ since these three coefficients
appear in the parameters
${\tilde \alpha}_i$ and
$p_i^2\!=\!q^2 \exp [ {\tilde u}_i(q^2) ]$ ($i\!=\!1,\!2$).

Concerning relations (\ref{tildeu21})-(\ref{det})
(for $M\!=\!2$), we can distinguish several cases 
\begin{itemize}
\item
When $(f_2\!-\!f_1^2)\!>\!0$, then:
${\tilde u}_i,\!{\tilde \alpha}_i$ are real ($i\!=\!1,\!2$),
${\tilde u}_1\!\not=\!{\tilde u}_2$ and 
$0\!<\!{\tilde \alpha}_i\!<\!1$.
\item
When $(f_2\!-\!f_1^2)\!<\!0$ and
$|f_3\!+\!f_1(2 f_1^2\!-\!3 f_2)|\!>\!2 \sqrt{ (f_1^2\!-\!f_2)^3 }$,
then: 
${\tilde u}_i,\!{\tilde \alpha}_i$ are real ($i\!=\!1,\!2$) and
${\tilde u}_1\!\not=\!{\tilde u}_2$.
\item
When $(f_2\!-\!f_1^2)\!<\!0$ and
$|f_3\!+\!f_1(2 f_1^2\!-\!3 f_2)|\!<\!2 \sqrt{ (f_1^2\!-\!f_2)^3 }$,
then: 
${\tilde u}_i$ are complex, ${\tilde \alpha}_i$
generally complex ($i\!=\!1,\!2$), and 
${\tilde u}_1\!\not=\!{\tilde u}_2$.
\item
When $(f_2\!-\!f_1^2)\!<\!0$ and
$|f_3\!+\!f_1(2 f_1^2\!-\!3 f_2)|\!=\!2 \sqrt{ (f_1^2\!-\!f_2)^3 }$
[or when $(f_2\!-\!f_1^2)\!=\!0$ and $f_3\!\not=\!f_1^3$], then:
the system of equations for ${\tilde u}_i$ and ${\tilde \alpha}_i$ is
not solvable, i.e., form (\ref{PAF1decomp}) is not valid,
the dPA (\ref{PAF11}) has a multiple (double) pole.
\item
When $(f_2\!-\!f_1^2)\!=\!0$ and $f_3\!=\!f_1^3$, then:
${\tilde u}_i,\!{\tilde \alpha}_i$ are real ($i\!=\!1,\!2$) and
${\tilde u}_1\!=\!{\tilde u}_2\!=\!f_1$.
\end{itemize}
Even when the parameters ${\tilde \alpha}_i$ and
${\tilde u}_i$ are complex [i.e., when `det' in
(\ref{tildeu21}) is negative], it can be checked directly
from (\ref{tildeu21}) that 
${\tilde \alpha}_2\!=\!({\tilde \alpha}_1)^{\ast}$ and
${\tilde u}_2\!=\!({\tilde u}_1)^{\ast}$, and thus
that the approximant is again real [note: 
$a((p_1^2)^{\ast})\!=\!a(p_1^2)^{\ast}$].

In QCD, presently available results
of perturbative calculations contain, for various observables
$S$ and in specific renormalization schemes, 
the coefficients $r_1(q^2)$ and $r_2(q^2)$ of (\ref{S}),
but not yet $r_3(q^2)$. Hence, the described
algorithm still cannot be applied for
$M\!=\!2$ for QCD observables, due to the fact
that $r_{2M\!-\!1}(q^2)\!\equiv\!r_3(q^2)$ are not yet known.
This, however, is in stark contrast with some QED observables
for which perturbative coefficients 
$r_3(q^2)$ have already been obtained.

The presented algorithm, although being a clear improvement
of the method of dPA's for perturbative series in gauge theories,
still has several deficiencies. 
One of them is that the obtained approximants probably cannot
discern in QCD, on the basis of a given TPS, the 
nonperturbative behavior originating from (ultraviolet und
infrared) renormalons -- see arguments in Refs.~\cite{SEK},
\cite{Gardi}-\cite{Brodskyetal} for the case of the usual (d)PA's
with which the presented modified dBGA approximants 
are closely related.
Another deficiency is that the algorithm
can be applied only in the cases when the available TPS
$S_n(q^2)$ of (\ref{Sn}) has odd $n\!=\!2M\!-\!1$, i.e.,
it cannot be applied in the case $n\!=\!2$ (which is
at present the case of many QCD observables). This is
so because the algorithm heavily relies on the decomposition
(\ref{PAF1decomp}) which is valid only for {\em diagonal}
PA's. This problematic restriction is present also in the
large-${\beta}_0$ limit, i.e., in the usual dPA approach.
Another problematic point is that 
the dBGA (\ref{dBGAdef})-(\ref{dBGAres})
becomes in the cases of $M\!\geq\!2$ also explicitly renormalization
scheme (RSch) dependent, because parameters 
${\tilde u}_i$ and ${\tilde \alpha}_i$ appearing in such dBGA's
also involve some of the ${\beta}_j$ ($j\!\geq\!2$) RGE
coefficients which are, in contrast to ${\beta}_0$ and ${\beta}_1$,
RSch-dependent.\footnote{
When all effects of ${\beta}_j$'s ($j\!\geq\!2$) are neglected,
changing the renormalization scheme is equivalent to
changing the renormalization scale.}
This contrasts with some other approaches. For example,
in the approach of the principle of minimal sensitivity (PMS)
\cite{Stevensonetal},
both the RScl- and RSch-independence
of the approximant are achieved via a local method, while
in the present approach the RScl-invariance is ensured via
a more global method and RSch-invariance (i.e., independence of
${\beta}_2,\!{\beta}_3,\!\ldots$) is not ensured at all.
It would definitely be instructive to compare the 
efficiency of the presented method (for $M\!=\!2$, for the
time being in QED only) with the PMS method, 
as well as with other methods, among them:
the effective charge (ECH) method \cite{Grunberg}, 
the Brodsky-Lepage-Mackenzie (BLM) approach
and its extensions \cite{BLM}--\cite{LMCF}
and \cite{Neubertetal}, 
a new approach \cite{Maxwell} based partly on the ECH methods,
and another new approach \cite{SoSh} using a method of analytic
continuation. It may be also useful to investigate which
classes of Feynman diagrams the 
presented algorithm approximately
sums up,\footnote{
Under the term ``approximately'' we understand here a
systematic approximation which would converge to
the sum of mentioned diagrams when $M\!\to\!\infty$.}
in analogy with the work \cite{Brodskyetal} for the
case of the dPA approach.

\vspace{0.5cm}

\noindent {\bf Abbreviations used frequently in the 
article\/:}\\ (d)BGA -- (diagonal) Baker-Gammel approximant; 
(d)PA -- (diagonal) Pad\'e approximant; 
RSch -- renormalization scheme; 
RScl -- renormalization scale;
TPS -- truncated perturbation series.

\vspace{0.5cm}

\noindent {\bf Acknowledgments\/:}

\noindent The author wishes to thank 
Professors D.~Schildknecht and R.~K\"ogerler
for offering him financial support of Bielefeld University
during the course of this work.


\begin{thebibliography}{99}

\bibitem{BakerMorris}
George A.~Baker, Jr. and Peter Graves-Morris,
Encyclopedia of Mathematics and Its Applications,
Pad\'e Approximants, Part I and II (Vol.~13 and 14), 
ed.~Gian-Carlo Rota (Addison-Wesley, 1981).

\bibitem{SEK}
M.~A.~Samuel, J.~Ellis and M.~Karliner, Phys. Rev. Lett. {\bf 74},
4380 (1995); J.~Ellis, E.~Gardi, M.~Karliner and M.~A.~Samuel,
Phys. Lett. B {\bf 366}, 268 (1996), and Phys. Rev. D {\bf 54}, 6986
(1996); I.~Jack, D.~R.~T.~Jones and M.~A.~Samuel, Phys. Lett. B {\bf 407}, 
143 (1997).

\bibitem{Fischer}
J.~Fischer, Int. J. Mod. Phys. A {\bf 12}, 3625 (1997).

\bibitem{Gardi}
E.~Gardi, Phys. Rev. D {\bf 56}, 68 (1997).

\bibitem{Brodskyetal}
S.~J.~Brodsky, J.~Ellis, E.~Gardi, M.~Karliner and M.~A.~Samuel,
Phys. Rev. D {\bf 56}, 6980 (1997).

\bibitem{Neubertetal}
Neubert, Phys. Rev. D {\bf 51}, 5942 (1995); see also:
M.~Beneke and V.~M.~Braun, Phys.~Lett. B {\bf 348}, 513 (1995).

\bibitem{QCDbeta}
T.~van Ritbergen, J.~A.~M.~Vermaseren and S.~A.~Larin,
Phys. Lett. B {\bf 400}, 379 (1997).

\bibitem{QEDbeta}
S.~G.~Gorishny, A.~L.~Kataev, S.~A.~Larin and L.~R.~Surguladze,
Phys. Lett. B {\bf 256}, 81 (1991); D.~J.~Broadhurst, A.~L.~Kataev and
O.~V.~Tarasov, Phys. Lett. B {\bf 298}, 445 (1993).

\bibitem{Cvetic}
G.~Cveti\v{c}, Bielefeld preprint BI-TP 97/52 (hep-ph/9711406).

\bibitem{Stevensonetal}
P.~M.~Stevenson, Phys. Rev. D {\bf 23}, 2961 (1981);
A.~L.~Kataev and V.~V.~Starshenko, Mod. Phys. Lett. A {\bf 10}, 
235 (1995).

\bibitem{Grunberg}
G.~Grunberg, Phys. Rev. D {\bf 29}, 2315 (1984).

\bibitem{BLM}
S.~J.~Brodsky, G.~P.~Lepage and P.~B.~Mackenzie,
Phys. Rev. D {\bf 28}, 228 (1983).

\bibitem{BLMext}
G.~Grunberg and A.~L.~Kataev, Phys. Lett. B {\bf 279}, 352 (1992);
S.~J.~Brodsky and H.~J.~Lu, Phys. Rev. D {\bf 51}, 3652 (1995);
S.~J.~Brodsky, G.~T.~Gabadadze, A.~L.~Kataev and H.~J.~Lu,
Phys. Lett. B {\bf 372}, 133 (1996);
J.~Rathsman, Phys. Rev. D {\bf 54}, 3420 (1996).

\bibitem{LMCF}
G.~P.~Lepage and P.~B.~Mackenzie, Phys. Rev. D {\bf 48}, 2250 (1993);
J.~C.~Collins and A.~Freund, Nucl. Phys. B {\bf 504}, 461 (1997).

\bibitem{Maxwell}
C.~J.~Maxwell, Phys. Lett B {\bf 409}, 450 (1997).

\bibitem{SoSh}
I.~L.~Solovtsov and D.~V.~Shirkov, hep-ph/9711251.


\end{thebibliography}
\end{document}